\documentclass[PRL,twocolumn,groupedaddress,nofootinbib,superscriptaddress]{revtex4-1}
\usepackage{graphicx}
\usepackage{dcolumn}
\usepackage{bm}
\usepackage{times}
\usepackage{epstopdf}
\usepackage[version=3]{mhchem} 
\usepackage{xcolor}

\usepackage{mathrsfs}
\usepackage{lipsum}
\usepackage{amsmath}
\usepackage{amsfonts}
\usepackage{array}
\usepackage{color}
\usepackage{ulem}

\newcommand{\degreesC}{\,$^{\circ}$C}

\newcommand{\cosnin}{CoSn$_{1-x}$In$_{x}$}

\begin{document}

\title{Flat-Band Itinerant Antiferromagnetism in the Kagome Metal \cosnin}

\author{B. C. Sales}
\email{salesbc@ornl.gov}
\affiliation{Materials Science and Technology Division, Oak Ridge National Laboratory, Oak Ridge, Tennessee 37831 USA}
\author{W. R. Meier}
\affiliation{Materials Science and Technology Division, Oak Ridge National Laboratory, Oak Ridge, Tennessee 37831 USA}
\author{D. S. Parker}
\affiliation{Materials Science and Technology Division, Oak Ridge National Laboratory, Oak Ridge, Tennessee 37831 USA}
\author{L. Yin}
\affiliation{Materials Science and Technology Division, Oak Ridge National Laboratory, Oak Ridge, Tennessee 37831 USA}
\author{J. Q. Yan}
\affiliation{Materials Science and Technology Division, Oak Ridge National Laboratory, Oak Ridge, Tennessee 37831 USA}
\author{A. F. May}
\affiliation{Materials Science and Technology Division, Oak Ridge National Laboratory, Oak Ridge, Tennessee 37831 USA}
\author{S. Calder}
\affiliation{Neutron Scattering Division, Oak ridge National Laboratory, Oak Ridge, Tennessee 37831, USA}
\author{A. A. Aczel}
\affiliation{Neutron Scattering Division, Oak ridge National Laboratory, Oak Ridge, Tennessee 37831, USA}
\author{Q. Zhang}
\affiliation{Neutron Scattering Division, Oak ridge National Laboratory, Oak Ridge, Tennessee 37831, USA}
\author{H. Li}
\affiliation{Materials Science and Technology Division, Oak Ridge National Laboratory, Oak Ridge, Tennessee 37831 USA}
\author{T. Yilmaz}
\affiliation{National Synchrotron Light Source II, Brookhaven National Laboratory, Upton, New York 11973, USA}
\author{E. Vescovo}
\affiliation{National Synchrotron Light Source II, Brookhaven National Laboratory, Upton, New York 11973, USA}
\author{H. Miao}
\affiliation{Materials Science and Technology Division, Oak Ridge National Laboratory, Oak Ridge, Tennessee 37831 USA}
\author{R. P. Hermann}
\affiliation{Materials Science and Technology Division, Oak Ridge National Laboratory, Oak Ridge, Tennessee 37831 USA}
\author{M. A. McGuire}
\email{mcguirema@ornl.gov \\  \\ Notice: This manuscript has been authored by UT-Battelle, LLC under Contract No. DE-AC05-00OR22725 with the U.S. Department of Energy. The United States Government retains and the publisher, by accepting the article for publication, acknowledges that the United States Government retains a non-exclusive, paid-up, irrevocable, world-wide license to publish or reproduce the published form of this manuscript, or allow others to do so, for United States Government purposes. The Department of Energy will provide public access to these results of federally sponsored research in accordance with the DOE Public Access Plan (http://energy.gov/downloads/doe-public-access-plan). }
\affiliation{Materials Science and Technology Division, Oak Ridge National Laboratory, Oak Ridge, Tennessee 37831 USA}

\begin{abstract}
Destructive interference of electron hopping on the frustrated kagome lattice generates Dirac nodes, saddle points, and flat bands in the electronic structure. The latter provides the narrow bands and a peak in the density of states that can generate correlated electron behavior when the Fermi level lies within them. In the kagome metal CoSn, this alignment is not realized, and the compound is a Pauli paramagnet. Here we show that replacing part of the tin with indium (\cosnin) moves the Fermi energy into the flat band region, with support from band structure calculations, heat capacity measurements, and angle resolved photoemission spectroscopy. The associated instability results in the emergence of itinerant antiferromagnetism with a Neel temperature up to 30\,K. Long range magnetic order is confirmed by neutron diffraction measurements, which indicate an ordered magnetic moment of 0.1-0.2 $\mu_B$ per Co (for x = 0.4). Thus, \cosnin\ provides a rare example of an itinerant antiferromagnet with a small ordered moment. This work provides clear evidence that flat bands arising from frustrated lattices in bulk crystals represent a viable route to new physics, evidenced here by the emergence of magnetic order upon introducing a non-magnetic dopant into a non-magnetic kagome metal.
\end{abstract}

\maketitle

The symmetry and connectivity of certain crystalline lattices can produce extremely narrow electronic bands in reciprocal space, i.e. flat bands. In such flat bands the electrons have a high density of states and strong many-body interactions that can lead to unusual non-Fermi liquid behavior and exotic ground states \cite{calugaru2021, sutherland1986, leykam2018artificial, regnault2021catalogue, mazin2014theoretical, sun2011nearly, tang2011high}. The most spectacular experimental verifications of these ideas are found in the Moir\'{e} materials such as twisted bilayer graphene structures, where unusual magnetism, superconductivity, and the quantum anomalous Hall effect have been reported \cite{cao2018unconventional, cao2021pauli, serlin2020intrinsic}.  Pyrochlore lattices in 3D and kagome, Lieb and dice lattices in 2D also generate flat electronic bands in simplified theoretical calculations \cite{calugaru2021, sutherland1986, leykam2018artificial, regnault2021catalogue, mazin2014theoretical, sun2011nearly, tang2011high}. The basic physical idea is that the destructive interference of electron hopping on these lattices results in electron localization and bands with little dispersion in momentum space.  Although bulk crystals are 3D, there are many quasi-2D layered compounds where each layer has the desired geometrical arrangement of atoms. A comprehensive approach to identify and classify all compounds that have relatively flat bands in the vicinity of the Fermi energy, and the topological character of these bands was recently reported using the Inorganic Crystal Structure Database (ICSD) and the Topological Quantum Chemistry website \cite{calugaru2021, regnault2021catalogue}. The thousands of potential flat band materials identified in the study were grouped further depending on whether the flat bands were due primarily to local atomic wavefunctions (like in many rare earth compounds) or to more extended Bloch-like wavefunctions. It is the latter materials that are predicted to host many exotic phenomena such as the fractional quantum Hall effect, unusual magnetism, and unconventional superconductivity \cite{regnault2021catalogue}.

Compounds with crystal structures containing kagome layers are relatively common, and include \ce{Fe3Sn2}, \ce{Fe3Sn}, FeSn, FeGe, CoSn, \ce{Co3Sn2S2}, the \ce{HfFe6Ge6} family, RhPb, NiIn, PtTl, and the recently reported compounds \ce{Ni3In}, and \ce{AV3Sb5} (A = K, Rb, or Cs) \cite{kang2020dirac, meier2020flat, ye2021flat, ortiz2019new, meier2019reorientation, sales2019electronic, sales2021tuning}.  However, evidence for the flat bands impacting the physical properties is rare, since they must be tuned to lie near the Fermi energy. In twisted bilayer graphene, tuning is accomplished with a gate voltage \cite{kim2017tunable, cao2018unconventional}, but in bulk crystals chemical doping or pressure can be used.

\begin{figure*}
\begin{center}
\includegraphics[width=6.0in]{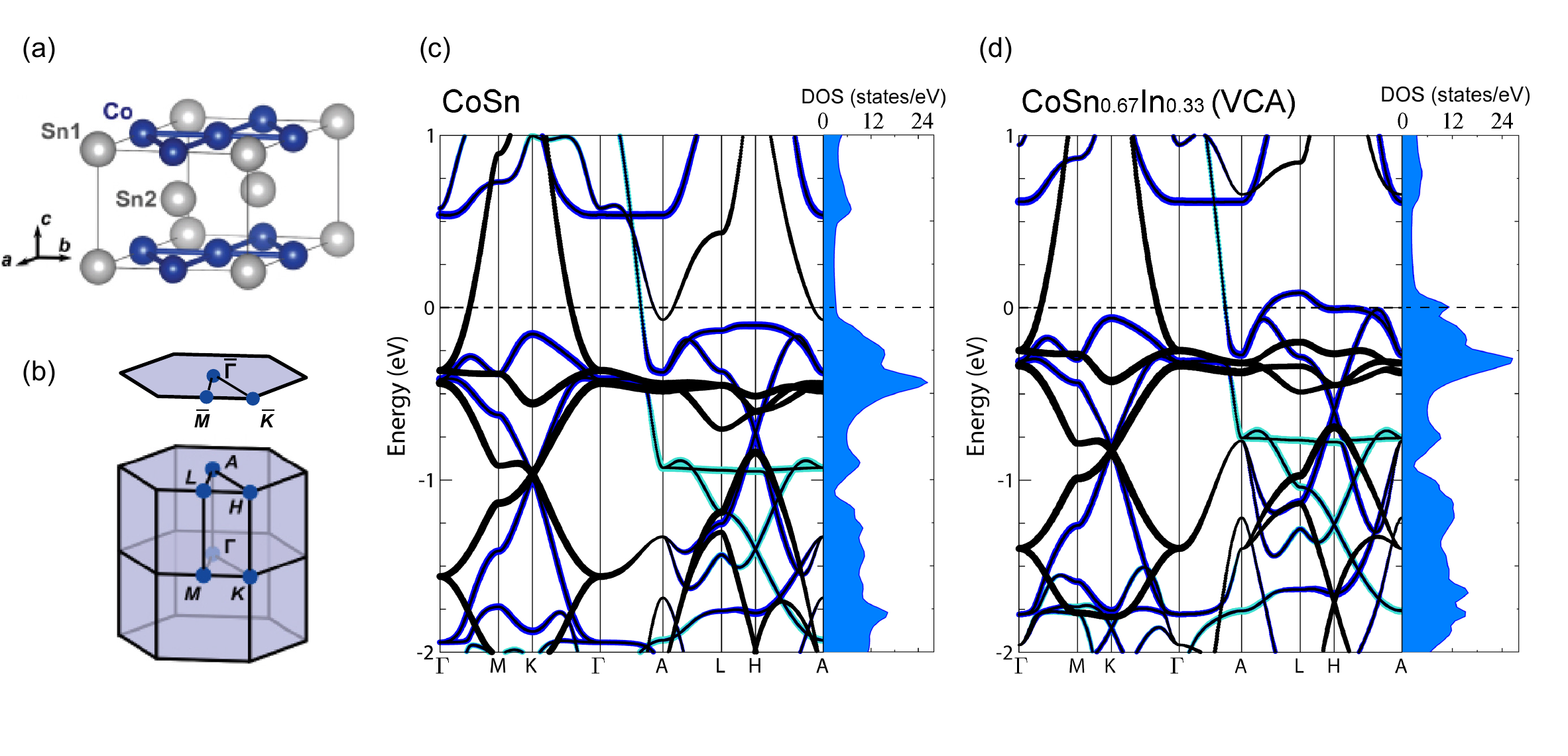}
\caption{\label{fig:bands}
Effect of In substitution on electronic structure of CoSn. (a) The crystal structure of CoSn. (b) The standard notation for the Brillioun zones for the hexagonal unit cell, with the surface Brillioun zone shown at the top. (c) The electronic structure of CoSn. (d) The electronic structure of CoSn$_{0.67}$In$_{0.33}$ calculated using the VCA. Bands in this energy range primarily originate from Co d orbitals, highlighted in the figure as dark blue for $d_{x2-y2}$ and $d_{xy}$, cyan for $d_{z2}$, and black for $d_{xz}$ and $d_{yz}$. The exception is the band producing the hole pocket near the A point in CoSn, which has strong Sn character. The DOS values are plotted per unit cell and there are three formula units per unit cell.
}
\end{center}
\end{figure*}

In CoSn, the flat bands are below the Fermi energy by about 100 meV, suggesting they can be accessed by hole doping. Previously we studied the effects of Fe or In doping on the properties of CoSn. Replacing 1-20\% of the Co with Fe resulted in a spin glass \cite{sales2021tuning}. Although adding Fe to CoSn did add holes and moved the flat bands closer to E$_F$, each Fe also appeared to carry a local magnetic moment of about 2 $\mu_B$ that dominated the magnetic response and obscured the underlying flat-band physics \cite{sales2021tuning}. Indium doping was limited in that study to concentrations only up to 20\%, and while showing no phase transition, indicated an approach toward a flat-band related instability with increasing In content.

In this Letter we report results obtained by replacing up to 40\% of the Sn in CoSn with In, which has one fewer valence electron than Sn. This substitution results in a movement of the Fermi level into the flat bands, while avoiding the complication of local moment magnetism seen with Fe doping. Magnetization measurements, neutron diffraction, and density functional theory calculations show that indeed a magnetic instability is induced. Angle resolved photoemission and heat capacity measurements confirm the flat-band-peak in the density of states is responsible for the transition. The ground state for high In doping (x\,$\approx$\,0.40) is identified as a long range ordered antiferromagnetic state by neutron diffraction, with a moment of 0.1-0.2 $\mu_B$ per Co and an ordering temperature of 24 K. A maximum Neel temperature of 30\,K is observed in this series by magnetization measurements. This work provides a clear example of a new correlated ground state emerging as the Fermi level approaches flat bands, evidenced by the appearance of magnetic order upon introducing a non-magnetic dopant into a non-magnetic kagome metal.

Details of the crystallography of \cosnin\ and growth of the samples used in this study can be found in the Supplemental Information \cite{supp}. The structure of CoSn is shown in Figure \ref{fig:bands}a. It contains a kagome net of Co (Wyckoff position 3f) and two Sn sites (1a, 2d), with three formula units in the hexagonal unit cell. Neutron diffraction and M\"{o}ssbauer spectroscopy show that In substitutes onto both Sn sites, with a preference for the 1a site over the 2a site apparent at high doping levels \cite{supp}.

We begin with simple electronic structure calculations that motivate our interest in \cosnin. These are summarized in Figure \ref{fig:bands}.  In CoSn (Fig. \ref{fig:bands}c), flat bands in the L-H-A plane are found about 100\,meV below the Fermi energy. The band structure for CoSn$_{0.67}$In$_{0.33}$ calculated using the virtual crystal approximation (VCA) is shown in Figure \ref{fig:bands}d. Indium substitution moves the Fermi energy down (akin to hole doping a semiconductor) and into the flat band region. The calculated density of states at E$_F$ is increased by more than a factor of three, from 3.1 to 11.2 eV$^{-1}$ per unit cell. In addition to moving the Fermi level into the flat band region, these calculations suggest indium substitution also produces some curvature to the flat band around the L point and pushes up the predominantly Sn band (the band producing the hole pocket near A in CoSn). Thus, while the effects of In substitution are more complicated than a simple rigid band shift, a clear enhancement in DOS at E$_F$ is seen, producing a tendency toward an electronic/magnetic instability arising from the flat bands.

\begin{figure*}
\begin{center}
\includegraphics[width=6.5in]{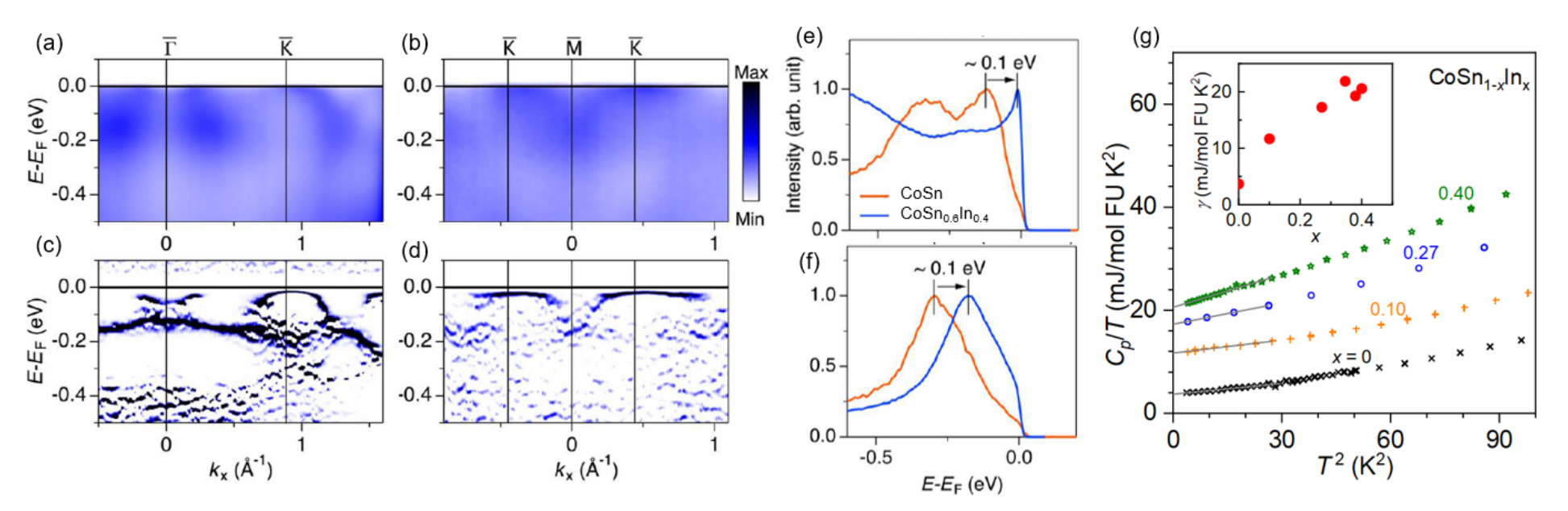}
\caption{\label{fig:arpes}
Evidence for the Fermi energy approaching the flat bands in \cosnin. (a,b) ARPES intensity plots from CoSn$_{0.6}$In$_{0.4}$ along the high symmetry $\overline{\Gamma}-\overline{K}$ and $\overline{M}-\overline{K}$ directions, showing flat bands near E$_F$. The curvature plots of (a) and (b) are shown in (c) and (d), respectively. Comparison of EDCs  for CoSn and CoSn$_{0.6}$In$_{0.4}$ at the $\overline{K}$ (e) and $\overline{\Gamma}$ (f) points show increased DOS near E$_F$ induced by In doping. (f) Low temperature specific heat data and fitted Sommerfeld coefficients $\gamma$ (inset), which increases in In content indicating higher DOS at E$_F$ in In doped crystals.
}
\end{center}
\end{figure*}

Figure \ref{fig:arpes} shows the electronic band structure of CoSn$_{0.6}$In$_{0.4}$ determined by angle-resolved photoemission spectroscopy (ARPES) measured at 15\,K. A map of the measured Fermi surface is provided in the Supplementary Material \cite{supp}. The ARPES intensity plots along the projected high symmetry directions $\overline{\Gamma}-\overline{K}$ and $\overline{M}-\overline{K}$  are shown in Fig. \ref{fig:arpes}a and \ref{fig:arpes}b, respectively. The corresponding curvature plots are shown in Fig. \ref{fig:arpes}(c,d). Along both directions, a partial flat band touching E$_F$ can be observed.  Figures \ref{fig:arpes}e and \ref{fig:arpes}f compare energy distribution curves (EDC) for CoSn and CoSn$_{0.6}$In$_{0.4}$ at the $\overline{\Gamma}$ and $\overline{K}$ points. While quantitative comparison of the ARPES data and DFT calculated band structures is likely not warranted, the $\sim$0.1 eV chemical potential shift is apparent in the data, pushing E$_F$ toward the flat band region and increasing the density of states at the Fermi level.  The change in the DOS at E$_F$ is also evidenced by low-temperature specific heat measurements, the results of which are shown in Figure \ref{fig:arpes}g. The composition dependence of the Sommerfeld coefficient $\gamma$ determined from low temperature heat capacity data is shown in the inset. The observed increase of $\gamma$ with x demonstrates that indium doping continually increases the DOS at E$_F$. Together the ARPES and heat capacity results provide clear experimental evidence that indium doping moves the Fermi level into the flat band region of the electronic structure in \cosnin, supporting the intuitive electron counting picture and the DFT band structure calculations. With this established, in the following we will show that the associated instability results in the emergence of flat band related magnetic order.

\begin{figure}
\begin{center}
\includegraphics[width=3.0in]{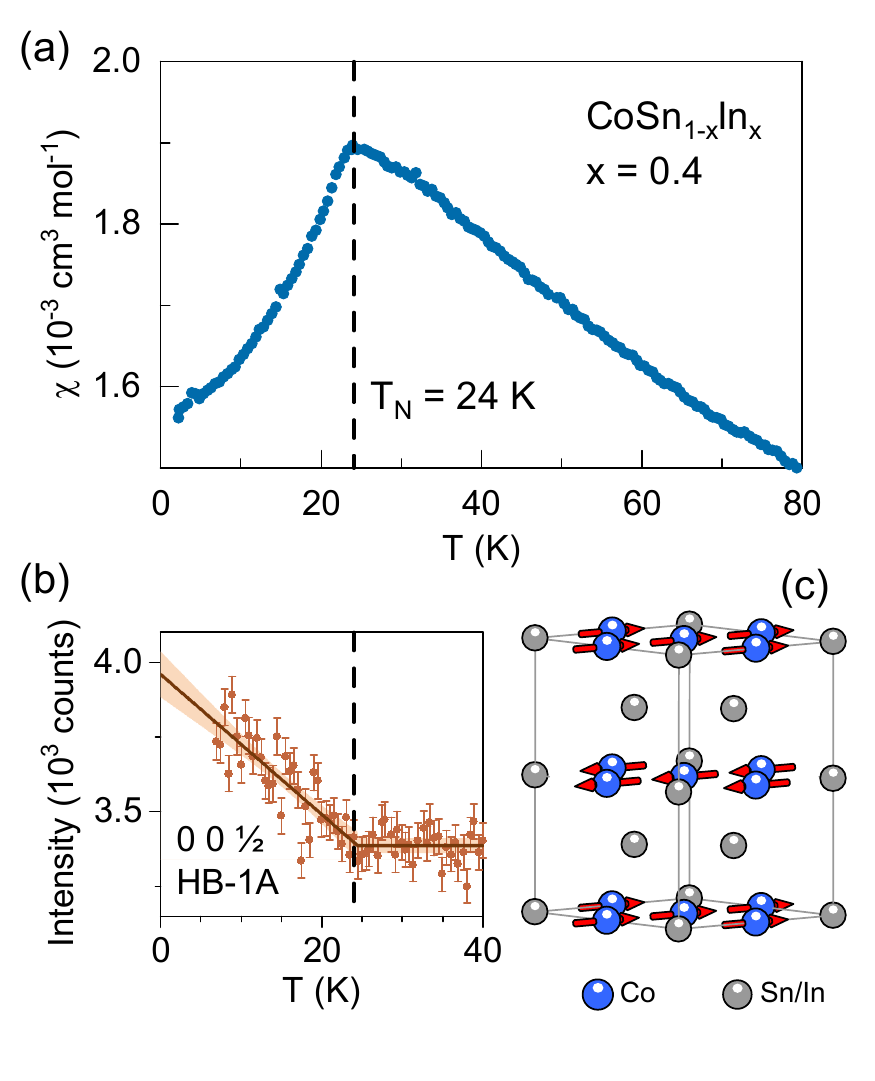}
\caption{\label{fig:magorder}
Emergent magnetic order in \cosnin. (a) Magnetic susceptibility from a CoSn$_{0.6}$In$_{0.4}$ powder (corrected to remove a contribution from a small cobalt impurity) showing a cusp indicative of antiferromagnetic order developing below $T_N$\,=\,24\,K. (b) Intensity of the 0 0 $\frac{1}{2}$ antiferromagnetic Bragg peak measured by neutron diffraction from the same powder, with an order parameter fit giving an onset of 24\,K (data from HB-1A). (c) A simple model magnetic structure consistent with the experimental data.
}
\end{center}
\end{figure}

As noted above, CoSn is a Pauli paramagnet and indium substitution up to 20\% has been shown to increase the overall magnetic moment \cite{sales2021tuning}. This is consistent with the electronic structure response just described, but no new ground state was observed at this modest indium content. Crystals with higher indium concentrations produced for the present study indeed reveal a transition to a long range magnetically ordered state. This is unambiguously demonstrated in Figure \ref{fig:magorder}. The magnetic susceptibility of a polycrystalline sample of CoSn$_{0.6}$In$_{0.4}$ shows a clear and sharp cusp at 24\,K (Fig. \ref{fig:magorder}a), and neutron diffraction from the same sample shows the onset of a magnetic Bragg peak at this same temperature. The Bragg peak is located at 0 0 $\frac{1}{2}$ position. Neutron diffraction patterns and more details of the measurements can be found in \cite{supp}. Since neutron diffraction measures the component of magnetization perpendicular to the scattering vector, and no other magnetic reflections are detected, the data suggests the ordered moments lie primarily in the ab plane, and are coupled antiferromagnetically along the c direction. This ordering is generally consistent with anisotropic magnetic susceptibility data shown below.

The simplest model magnetic structure consistent with the data is A-type antiferromagnetic order, as shown in Figure \ref{fig:magorder}c, although canting of the moments within the plane, which has been suggested for FeSn \cite{haggstrom1975studies}, cannot be ruled out. From the magnitude of the 0\,0\,$\frac{1}{2}$ peak in the full diffraction patterns, and assuming this magnetic structure, the ordered moment is refined to be 0.1-0.2 $\mu_B$ per Co, near the minimum limit for the neutron diffraction techniques used here. Diffraction data from a sample with less indium (CoSn$_{0.62}$In$_{0.38}$) suggested an even smaller moment, with the magnetic peak barely detectable. The smallness of the ordered moments is likely responsible for the absence of clear evidence of the transition in resistivity or heat capacity data, although the gradual development of magnetic correlations well above T$_N$ may also diminish features in those measurements at the transition to long range order \cite{supp}. Together magnetization and neutron scattering data show that CoSn$_{0.6}$In$_{0.4}$ undergoes a transition to an antiferromagnetically ordered state at $T_N$\,=\,24\,K with small ordered moments consistent with itinerant magnetism.

\begin{figure}
\begin{center}
\includegraphics[width=3.0in]{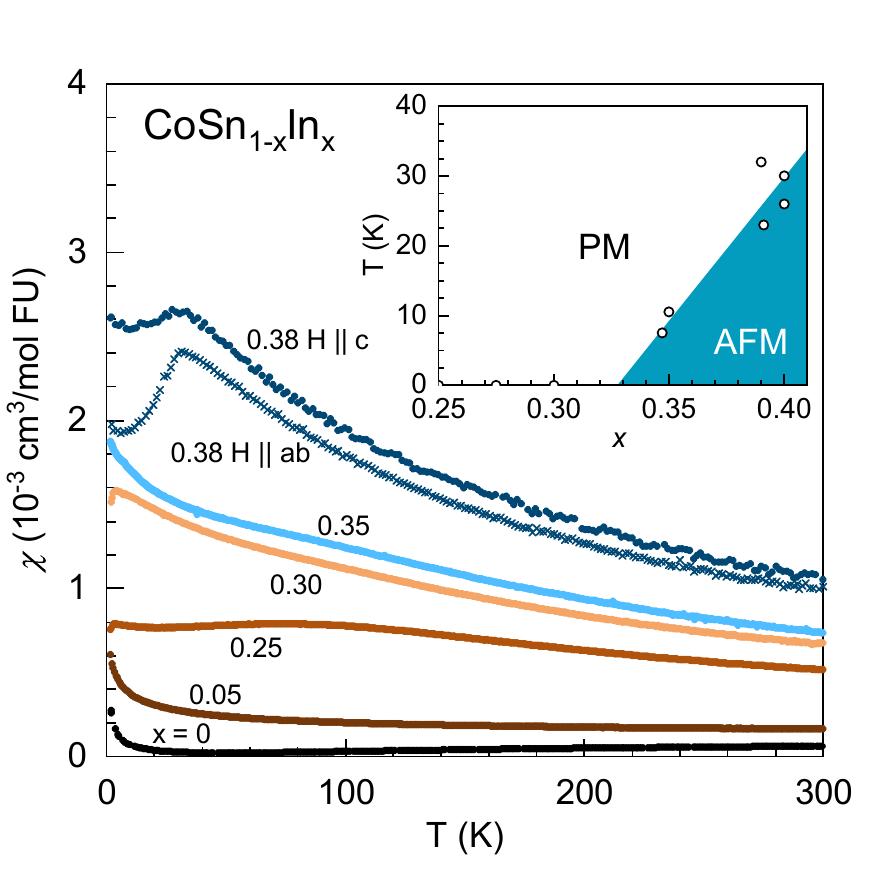}
\caption{\label{fig:PD}
Magnetic phase diagram of \cosnin. The temperature dependence of the magnetic susceptibility of several polycrystalline and single crystal \cosnin\ samples with varying values of x, measured in a field of 1\,kOe. Formation of AFM order is evidenced by a subtle cusp near 7.5\,K for x\,=\,0.35 and the clear maximum and anisotropic behavior for x\,=\,0.38. The downturns in $\chi$ at low T for x\,=\,0.25 and 0.30 are attributed to a superconducting Sn-In alloy present in residual flux on the crystals. A phase diagram constructed from these and other measurements is shown in the inset. The solubility limit is near $x=0.40$. Points at T\,=\,0 indicate no magnetic transition was detected down to 2\,K.
}
\end{center}
\end{figure}

The evolution of the magnetism in \cosnin, from Pauli paramagnetism at $x$\,=\,0 to antiferromagnetic order near $x$\,=\,0.4, is summarized in Figure \ref{fig:PD}. For low In content, Pauli paramagnetic like behavior is generally observed in the temperature dependence of the magnetic susceptibility $\chi$. When x exceeds about 0.25, Curie-Weiss like behavior emerges at higher temperatures. Note that the slight downturn in $\chi$ at the lowest temperatures seen for x = 0.25 and 0.30 is attributed to a superconducting transition in a Sn-In alloy present in residual flux on the crystals. For samples with x = 0.35 a subtle cusp in $\chi$ is observed at 8-10 K. When x approaches 0.40, a clear indication of the antiferromagnetic order described above is seen. We note that the $x$ values used here were derived from energy dispersive spectroscopy measurements and the trend in lattice parameters with composition described in \cite{supp}, and we ascribe an uncertainty in $x$ of about $\pm$0.02.

The magnetic phase diagram of \cosnin\ is shown in the inset of Figure \ref{fig:PD}, based on magnetization measurements like those shown in Figures \ref{fig:magorder} and \ref{fig:PD}. This is constructed from measurements on many samples, including both single crystal and polycrystalline forms. Magnetic order emerges at $0.3<x<0.35$ and the highest measured ordering temperatures are 30\,K for a polycrystalline sample with $x \approx 0.40$.

Data is shown in Figure \ref{fig:PD} for a single crystal of CoSn$_{0.62}$In$_{0.38}$ with magnetic fields along c and in the ab plane. The temperature dependence suggests an antiferromagnetic ground state with moments primarily in the ab-plane, consistent with the neutron diffraction discussed above. This sample has a Neel temperature of 30 K. Well above T$_N$ the susceptibility is accurately described by a Curie-Weiss law, C/(T - $\theta$). While Curie Weiss behavior is often associated with local magnetic moments, it is not uncommon to find this behavior in itinerant magnet systems as well, likely related to fluctuations, and even when in compounds with no “magnetic” elements \cite{svanidze2015itinerant}. This has been observed in the itinerant antiferromagnet TiAu, with an effective moment of 0.4 $\mu_B$/atom \cite{svanidze2015itinerant}, as well as the itinerant ferromagnet \ce{Sc3In} with an effective moment of 0.5 $\mu_B$/atom \cite{matthias1961ferromagnetism}. Fitting the data for CoSn$_{0.62}$In$_{0.38}$ gives $\theta$ = -140 K, and an effective moment $\mu_{eff}$ = 1.92 $\mu_B$ per Co. The Weiss temperature is about 5 times larger than $T_N$ suggesting significant frustration  and, interestingly, $\mu_{eff}$ is close to the S = 1/2 value of 1.72 $\mu_B$.

Finally, we return to the first principles calculations to examine the nature of the magnetism in \cosnin. These all-electron calculations were performed within the Generalized Gradient Approximation, with details provided in \cite{supp}. The In and Sn distribution was modeled in two ways, the VCA, as used for the band structures shown in Figure \ref{fig:bands}, and by using ordered structures with the two Sn sites either fully occupied by Sn or by In. The calculations showed a preference for In to occupy the Sn1 site, consistent with experiment. The total energy per unit cell is over 100 meV lower when In is substituted for Sn1 than it is when In is substituted for Sn2. Thus, calculations were done for CoSn$_{0.67}$In$_{0.33}$, with In fully replacing Sn1 in the ordered cell model, and with a 2:1 mixture of In and Sn on both sites in the VCA.

We first note that our DFT calculations produce non-magnetic solutions for pure CoSn. For CoSn$_{0.67}$In$_{0.33}$ on the other hand, total energy calculations for both models provide strong evidence of a magnetic instability. Energies of ferromagnetic (FM) and A-type antiferromagnetic (AF) structures fall well below the those from the non-magnetic calculations (see table in \cite{supp}). This emergence of magnetism with In doping is consistent with experimental observations and is a direct consequence of the movement of the Fermi level into flatter, higher density-of-states bands below the CoSn Fermi level. A small ordered moment is found in the calculations (0.3-0.4\,$\mu_B$ per Co). In conflict with the experimental data, the calculations predict FM order to be favored over AF by about 0.5 meV per Co in the ordered model and 3 meV per Co in the VCA. We do note that this calculated energy difference (i.e. AF – FM) is relatively small, especially for the ordered model, and strongly dependent on the number of k-points used. As many as 10,000 k-points in the full Brillouin zone were required to obtain sufficient convergence. This result indicates that the ground state is sensitive to subtle changes in the electronic and lattice structures.  This may in fact also reflect the importance of flat bands in CoSn, which present a challenge to first principles calculations. We also note that up to this point, correlations have been neglected in the theory. To determine how correlations may affect the magnetic order, calculations were performed with GGA+U and a moderate value of U\,=\,2\,eV on Co. This indeed changes the calculated ground state to the experimentally observed AF state, falling some 11 meV per Co below the FM state. While this calculation results in a too-large cobalt moment of 1.2\,$\mu_B$, it is nonetheless suggestive of the potential role of electronic correlations in the observed magnetic behavior.

The calculations also provide insight into the itinerant nature of the magnetism in \cosnin. Normally, itinerant behavior, as opposed to local moment behavior, is theoretically established by the inability to converge magnetic states consistent with the magnetic pattern chosen during initialization of the calculations \cite{shanavas2015itinerant}. That is, during convergence the system spontaneously chooses an alternative arrangement, potentially the ground state. Such behavior generally indicates that the magnetic moments involved do not possess individual or ``local'' character, where the magnitude of the individual moments is largely independent of the particular ordering pattern modeled. See Ref. \citenum{shanavas2015itinerant} for additional details. To test this, we performed a calculation (using the ordered CoSn$_{0.67}$In$_{0.33}$ cell) starting from a ferrimagnetic structure with one of the three Co moments antialigned to the other two. When initialized in this way, the calculation converged to the ferromagnetic solution, with the single Co moment initialized opposite to the other two spontaneously flipping to coalign with the others. Thus, the first principles calculations indicate alloying In into CoSn induces generally itinerant magnetism, as found experimentally.

In summary, using \cosnin\ we have demonstrated the emergence of magnetic order in a non-magnetic kagome metal by using a non-magnetic dopant to tune the Fermi energy. The movement of E$_F$ to energies with high DOS associated with flat bands is confirmed by first principles calculations, ARPES, and heat capacity measurements. The resulting magnetic order is identified in magnetization measurements and neutron diffraction, and A-type antiferromagnetic order with small moments lying within the kagome planes is the simplest model consistent with the data. The experimental data and theoretical calculations suggest \cosnin\ is a relatively rare example of an antiferromagnet with small itinerant moments. This work provides clear evidence of a new ground state associated with flat bands emerging in a bulk crystalline material. This should motivate further detailed studies of magnetism in this system, but more importantly the exploration of other single-crystalline bulk materials where flat bands may be exploited to reveal emergent states through chemical control of the Fermi energy.

\section*{Acknowledgements}
We thank Joe Paddison for helpful discussions. This work was funded by the Department of Energy, Office of Science, Basic Energy Sciences, Materials Sciences and Engineering Division. Work at the Oak Ridge National Laboratory Spallation Neutron Source and the High Flux Isotope Reactor was supported by U.S. DOE, Office of Science, Basic Energy Sciences, Scientific User Facilities Division.


\begin{thebibliography}{26}%
\makeatletter
\providecommand \@ifxundefined [1]{%
 \@ifx{#1\undefined}
}%
\providecommand \@ifnum [1]{%
 \ifnum #1\expandafter \@firstoftwo
 \else \expandafter \@secondoftwo
 \fi
}%
\providecommand \@ifx [1]{%
 \ifx #1\expandafter \@firstoftwo
 \else \expandafter \@secondoftwo
 \fi
}%
\providecommand \natexlab [1]{#1}%
\providecommand \enquote  [1]{``#1''}%
\providecommand \bibnamefont  [1]{#1}%
\providecommand \bibfnamefont [1]{#1}%
\providecommand \citenamefont [1]{#1}%
\providecommand \href@noop [0]{\@secondoftwo}%
\providecommand \href [0]{\begingroup \@sanitize@url \@href}%
\providecommand \@href[1]{\@@startlink{#1}\@@href}%
\providecommand \@@href[1]{\endgroup#1\@@endlink}%
\providecommand \@sanitize@url [0]{\catcode `\\12\catcode `\$12\catcode
  `\&12\catcode `\#12\catcode `\^12\catcode `\_12\catcode `\%12\relax}%
\providecommand \@@startlink[1]{}%
\providecommand \@@endlink[0]{}%
\providecommand \url  [0]{\begingroup\@sanitize@url \@url }%
\providecommand \@url [1]{\endgroup\@href {#1}{\urlprefix }}%
\providecommand \urlprefix  [0]{URL }%
\providecommand \Eprint [0]{\href }%
\providecommand \doibase [0]{http://dx.doi.org/}%
\providecommand \selectlanguage [0]{\@gobble}%
\providecommand \bibinfo  [0]{\@secondoftwo}%
\providecommand \bibfield  [0]{\@secondoftwo}%
\providecommand \translation [1]{[#1]}%
\providecommand \BibitemOpen [0]{}%
\providecommand \bibitemStop [0]{}%
\providecommand \bibitemNoStop [0]{.\EOS\space}%
\providecommand \EOS [0]{\spacefactor3000\relax}%
\providecommand \BibitemShut  [1]{\csname bibitem#1\endcsname}%
\let\auto@bib@innerbib\@empty
\bibitem [{\citenamefont {C{\u{a}}lug{\u{a}}ru}\ \emph
  {et~al.}(2021)\citenamefont {C{\u{a}}lug{\u{a}}ru}, \citenamefont {Chew},
  \citenamefont {Elcoro}, \citenamefont {Regnault}, \citenamefont {Song},\ and\
  \citenamefont {Bernevig}}]{calugaru2021}%
  \BibitemOpen
  \bibfield  {author} {\bibinfo {author} {\bibfnamefont {D.}~\bibnamefont
  {C{\u{a}}lug{\u{a}}ru}}, \bibinfo {author} {\bibfnamefont {A.}~\bibnamefont
  {Chew}}, \bibinfo {author} {\bibfnamefont {L.}~\bibnamefont {Elcoro}},
  \bibinfo {author} {\bibfnamefont {N.}~\bibnamefont {Regnault}}, \bibinfo
  {author} {\bibfnamefont {Z.~D.}\ \bibnamefont {Song}}, \ and\ \bibinfo
  {author} {\bibfnamefont {B.~A.}\ \bibnamefont {Bernevig}},\ }\href@noop {}
  {\enquote {\bibinfo {title} {General construction and topological
  classification of all magnetic and non-magnetic flat bands},}\ } (\bibinfo
  {year} {2021}),\ \Eprint {http://arxiv.org/abs/arXiv: 2106.05272v1} {arXiv:
  2106.05272v1} \BibitemShut {NoStop}%
\bibitem [{\citenamefont {Sutherland}(1986)}]{sutherland1986}%
  \BibitemOpen
  \bibfield  {author} {\bibinfo {author} {\bibfnamefont {B.}~\bibnamefont
  {Sutherland}},\ }\href@noop {} {\bibfield  {journal} {\bibinfo  {journal}
  {Phys. Rev. B}\ }\textbf {\bibinfo {volume} {34}},\ \bibinfo {pages} {5208}
  (\bibinfo {year} {1986})}\BibitemShut {NoStop}%
\bibitem [{\citenamefont {Leykam}\ \emph {et~al.}(2018)\citenamefont {Leykam},
  \citenamefont {Andreanov},\ and\ \citenamefont
  {Flach}}]{leykam2018artificial}%
  \BibitemOpen
  \bibfield  {author} {\bibinfo {author} {\bibfnamefont {D.}~\bibnamefont
  {Leykam}}, \bibinfo {author} {\bibfnamefont {A.}~\bibnamefont {Andreanov}}, \
  and\ \bibinfo {author} {\bibfnamefont {S.}~\bibnamefont {Flach}},\
  }\href@noop {} {\bibfield  {journal} {\bibinfo  {journal} {ADV PHYS-X}\
  }\textbf {\bibinfo {volume} {3}},\ \bibinfo {pages} {1473052} (\bibinfo
  {year} {2018})}\BibitemShut {NoStop}%
\bibitem [{\citenamefont {Regnault}\ \emph {et~al.}(2021)\citenamefont
  {Regnault}, \citenamefont {Xu}, \citenamefont {Li}, \citenamefont {Ma},
  \citenamefont {Jovanovic}, \citenamefont {Yazdani}, \citenamefont {Parkin},
  \citenamefont {Felser}, \citenamefont {Schoop}, \citenamefont {Ong} \emph
  {et~al.}}]{regnault2021catalogue}%
  \BibitemOpen
  \bibfield  {author} {\bibinfo {author} {\bibfnamefont {N.}~\bibnamefont
  {Regnault}}, \bibinfo {author} {\bibfnamefont {Y.}~\bibnamefont {Xu}},
  \bibinfo {author} {\bibfnamefont {M.-R.}\ \bibnamefont {Li}}, \bibinfo
  {author} {\bibfnamefont {D.-S.}\ \bibnamefont {Ma}}, \bibinfo {author}
  {\bibfnamefont {M.}~\bibnamefont {Jovanovic}}, \bibinfo {author}
  {\bibfnamefont {A.}~\bibnamefont {Yazdani}}, \bibinfo {author} {\bibfnamefont
  {S.~S.}\ \bibnamefont {Parkin}}, \bibinfo {author} {\bibfnamefont
  {C.}~\bibnamefont {Felser}}, \bibinfo {author} {\bibfnamefont {L.~M.}\
  \bibnamefont {Schoop}}, \bibinfo {author} {\bibfnamefont {N.~P.}\
  \bibnamefont {Ong}},  \emph {et~al.},\ }\href@noop {} {\bibfield  {journal}
  {\bibinfo  {journal} {arXiv preprint arXiv:2106.05287}\ } (\bibinfo {year}
  {2021})}\BibitemShut {NoStop}%
\bibitem [{\citenamefont {Mazin}\ \emph {et~al.}(2014)\citenamefont {Mazin},
  \citenamefont {Jeschke}, \citenamefont {Lechermann}, \citenamefont {Lee},
  \citenamefont {Fink}, \citenamefont {Thomale},\ and\ \citenamefont
  {Valent{\'\i}}}]{mazin2014theoretical}%
  \BibitemOpen
  \bibfield  {author} {\bibinfo {author} {\bibfnamefont {I.}~\bibnamefont
  {Mazin}}, \bibinfo {author} {\bibfnamefont {H.~O.}\ \bibnamefont {Jeschke}},
  \bibinfo {author} {\bibfnamefont {F.}~\bibnamefont {Lechermann}}, \bibinfo
  {author} {\bibfnamefont {H.}~\bibnamefont {Lee}}, \bibinfo {author}
  {\bibfnamefont {M.}~\bibnamefont {Fink}}, \bibinfo {author} {\bibfnamefont
  {R.}~\bibnamefont {Thomale}}, \ and\ \bibinfo {author} {\bibfnamefont
  {R.}~\bibnamefont {Valent{\'\i}}},\ }\href@noop {} {\bibfield  {journal}
  {\bibinfo  {journal} {Nat. Commun.}\ }\textbf {\bibinfo {volume} {5}},\
  \bibinfo {pages} {1} (\bibinfo {year} {2014})}\BibitemShut {NoStop}%
\bibitem [{\citenamefont {Sun}\ \emph {et~al.}(2011)\citenamefont {Sun},
  \citenamefont {Gu}, \citenamefont {Katsura},\ and\ \citenamefont
  {Sarma}}]{sun2011nearly}%
  \BibitemOpen
  \bibfield  {author} {\bibinfo {author} {\bibfnamefont {K.}~\bibnamefont
  {Sun}}, \bibinfo {author} {\bibfnamefont {Z.}~\bibnamefont {Gu}}, \bibinfo
  {author} {\bibfnamefont {H.}~\bibnamefont {Katsura}}, \ and\ \bibinfo
  {author} {\bibfnamefont {S.~D.}\ \bibnamefont {Sarma}},\ }\href@noop {}
  {\bibfield  {journal} {\bibinfo  {journal} {Phys. Rev. Lett.}\ }\textbf
  {\bibinfo {volume} {106}},\ \bibinfo {pages} {236803} (\bibinfo {year}
  {2011})}\BibitemShut {NoStop}%
\bibitem [{\citenamefont {Tang}\ \emph {et~al.}(2011)\citenamefont {Tang},
  \citenamefont {Mei},\ and\ \citenamefont {Wen}}]{tang2011high}%
  \BibitemOpen
  \bibfield  {author} {\bibinfo {author} {\bibfnamefont {E.}~\bibnamefont
  {Tang}}, \bibinfo {author} {\bibfnamefont {J.-W.}\ \bibnamefont {Mei}}, \
  and\ \bibinfo {author} {\bibfnamefont {X.-G.}\ \bibnamefont {Wen}},\
  }\href@noop {} {\bibfield  {journal} {\bibinfo  {journal} {Phys. Rev. Lett.}\
  }\textbf {\bibinfo {volume} {106}},\ \bibinfo {pages} {236802} (\bibinfo
  {year} {2011})}\BibitemShut {NoStop}%
\bibitem [{\citenamefont {Cao}\ \emph {et~al.}(2018)\citenamefont {Cao},
  \citenamefont {Fatemi}, \citenamefont {Fang}, \citenamefont {Watanabe},
  \citenamefont {Taniguchi}, \citenamefont {Kaxiras},\ and\ \citenamefont
  {Jarillo-Herrero}}]{cao2018unconventional}%
  \BibitemOpen
  \bibfield  {author} {\bibinfo {author} {\bibfnamefont {Y.}~\bibnamefont
  {Cao}}, \bibinfo {author} {\bibfnamefont {V.}~\bibnamefont {Fatemi}},
  \bibinfo {author} {\bibfnamefont {S.}~\bibnamefont {Fang}}, \bibinfo {author}
  {\bibfnamefont {K.}~\bibnamefont {Watanabe}}, \bibinfo {author}
  {\bibfnamefont {T.}~\bibnamefont {Taniguchi}}, \bibinfo {author}
  {\bibfnamefont {E.}~\bibnamefont {Kaxiras}}, \ and\ \bibinfo {author}
  {\bibfnamefont {P.}~\bibnamefont {Jarillo-Herrero}},\ }\href@noop {}
  {\bibfield  {journal} {\bibinfo  {journal} {Nature}\ }\textbf {\bibinfo
  {volume} {556}},\ \bibinfo {pages} {43} (\bibinfo {year} {2018})}\BibitemShut
  {NoStop}%
\bibitem [{\citenamefont {Cao}\ \emph {et~al.}(2021)\citenamefont {Cao},
  \citenamefont {Park}, \citenamefont {Watanabe}, \citenamefont {Taniguchi},\
  and\ \citenamefont {Jarillo-Herrero}}]{cao2021pauli}%
  \BibitemOpen
  \bibfield  {author} {\bibinfo {author} {\bibfnamefont {Y.}~\bibnamefont
  {Cao}}, \bibinfo {author} {\bibfnamefont {J.~M.}\ \bibnamefont {Park}},
  \bibinfo {author} {\bibfnamefont {K.}~\bibnamefont {Watanabe}}, \bibinfo
  {author} {\bibfnamefont {T.}~\bibnamefont {Taniguchi}}, \ and\ \bibinfo
  {author} {\bibfnamefont {P.}~\bibnamefont {Jarillo-Herrero}},\ }\href@noop {}
  {\bibfield  {journal} {\bibinfo  {journal} {Nature}\ }\textbf {\bibinfo
  {volume} {595}},\ \bibinfo {pages} {526} (\bibinfo {year}
  {2021})}\BibitemShut {NoStop}%
\bibitem [{\citenamefont {Serlin}\ \emph {et~al.}(2020)\citenamefont {Serlin},
  \citenamefont {Tschirhart}, \citenamefont {Polshyn}, \citenamefont {Zhang},
  \citenamefont {Zhu}, \citenamefont {Watanabe}, \citenamefont {Taniguchi},
  \citenamefont {Balents},\ and\ \citenamefont {Young}}]{serlin2020intrinsic}%
  \BibitemOpen
  \bibfield  {author} {\bibinfo {author} {\bibfnamefont {M.}~\bibnamefont
  {Serlin}}, \bibinfo {author} {\bibfnamefont {C.}~\bibnamefont {Tschirhart}},
  \bibinfo {author} {\bibfnamefont {H.}~\bibnamefont {Polshyn}}, \bibinfo
  {author} {\bibfnamefont {Y.}~\bibnamefont {Zhang}}, \bibinfo {author}
  {\bibfnamefont {J.}~\bibnamefont {Zhu}}, \bibinfo {author} {\bibfnamefont
  {K.}~\bibnamefont {Watanabe}}, \bibinfo {author} {\bibfnamefont
  {T.}~\bibnamefont {Taniguchi}}, \bibinfo {author} {\bibfnamefont
  {L.}~\bibnamefont {Balents}}, \ and\ \bibinfo {author} {\bibfnamefont
  {A.}~\bibnamefont {Young}},\ }\href@noop {} {\bibfield  {journal} {\bibinfo
  {journal} {Science}\ }\textbf {\bibinfo {volume} {367}},\ \bibinfo {pages}
  {900} (\bibinfo {year} {2020})}\BibitemShut {NoStop}%
\bibitem [{\citenamefont {Kang}\ \emph {et~al.}(2020)\citenamefont {Kang},
  \citenamefont {Ye}, \citenamefont {Fang}, \citenamefont {You}, \citenamefont
  {Levitan}, \citenamefont {Han}, \citenamefont {Facio}, \citenamefont
  {Jozwiak}, \citenamefont {Bostwick}, \citenamefont {Rotenberg} \emph
  {et~al.}}]{kang2020dirac}%
  \BibitemOpen
  \bibfield  {author} {\bibinfo {author} {\bibfnamefont {M.}~\bibnamefont
  {Kang}}, \bibinfo {author} {\bibfnamefont {L.}~\bibnamefont {Ye}}, \bibinfo
  {author} {\bibfnamefont {S.}~\bibnamefont {Fang}}, \bibinfo {author}
  {\bibfnamefont {J.-S.}\ \bibnamefont {You}}, \bibinfo {author} {\bibfnamefont
  {A.}~\bibnamefont {Levitan}}, \bibinfo {author} {\bibfnamefont
  {M.}~\bibnamefont {Han}}, \bibinfo {author} {\bibfnamefont {J.~I.}\
  \bibnamefont {Facio}}, \bibinfo {author} {\bibfnamefont {C.}~\bibnamefont
  {Jozwiak}}, \bibinfo {author} {\bibfnamefont {A.}~\bibnamefont {Bostwick}},
  \bibinfo {author} {\bibfnamefont {E.}~\bibnamefont {Rotenberg}},  \emph
  {et~al.},\ }\href@noop {} {\bibfield  {journal} {\bibinfo  {journal} {Nat.
  Mater.}\ }\textbf {\bibinfo {volume} {19}},\ \bibinfo {pages} {163} (\bibinfo
  {year} {2020})}\BibitemShut {NoStop}%
\bibitem [{\citenamefont {Meier}\ \emph {et~al.}(2020)\citenamefont {Meier},
  \citenamefont {Du}, \citenamefont {Okamoto}, \citenamefont {Mohanta},
  \citenamefont {May}, \citenamefont {McGuire}, \citenamefont {Bridges},
  \citenamefont {Samolyuk},\ and\ \citenamefont {Sales}}]{meier2020flat}%
  \BibitemOpen
  \bibfield  {author} {\bibinfo {author} {\bibfnamefont {W.~R.}\ \bibnamefont
  {Meier}}, \bibinfo {author} {\bibfnamefont {M.-H.}\ \bibnamefont {Du}},
  \bibinfo {author} {\bibfnamefont {S.}~\bibnamefont {Okamoto}}, \bibinfo
  {author} {\bibfnamefont {N.}~\bibnamefont {Mohanta}}, \bibinfo {author}
  {\bibfnamefont {A.~F.}\ \bibnamefont {May}}, \bibinfo {author} {\bibfnamefont
  {M.~A.}\ \bibnamefont {McGuire}}, \bibinfo {author} {\bibfnamefont {C.~A.}\
  \bibnamefont {Bridges}}, \bibinfo {author} {\bibfnamefont {G.~D.}\
  \bibnamefont {Samolyuk}}, \ and\ \bibinfo {author} {\bibfnamefont {B.~C.}\
  \bibnamefont {Sales}},\ }\href@noop {} {\bibfield  {journal} {\bibinfo
  {journal} {Phys. Rev. B}\ }\textbf {\bibinfo {volume} {102}},\ \bibinfo
  {pages} {075148} (\bibinfo {year} {2020})}\BibitemShut {NoStop}%
\bibitem [{\citenamefont {Ye}\ \emph {et~al.}(2021)\citenamefont {Ye},
  \citenamefont {Fang}, \citenamefont {Kang}, \citenamefont {Kaufmann},
  \citenamefont {Lee}, \citenamefont {Denlinger}, \citenamefont {Jozwiak},
  \citenamefont {Bostwick}, \citenamefont {Rotenberg}, \citenamefont {Kaxiras}
  \emph {et~al.}}]{ye2021flat}%
  \BibitemOpen
  \bibfield  {author} {\bibinfo {author} {\bibfnamefont {L.}~\bibnamefont
  {Ye}}, \bibinfo {author} {\bibfnamefont {S.}~\bibnamefont {Fang}}, \bibinfo
  {author} {\bibfnamefont {M.~G.}\ \bibnamefont {Kang}}, \bibinfo {author}
  {\bibfnamefont {J.}~\bibnamefont {Kaufmann}}, \bibinfo {author}
  {\bibfnamefont {Y.}~\bibnamefont {Lee}}, \bibinfo {author} {\bibfnamefont
  {J.}~\bibnamefont {Denlinger}}, \bibinfo {author} {\bibfnamefont
  {C.}~\bibnamefont {Jozwiak}}, \bibinfo {author} {\bibfnamefont
  {A.}~\bibnamefont {Bostwick}}, \bibinfo {author} {\bibfnamefont
  {E.}~\bibnamefont {Rotenberg}}, \bibinfo {author} {\bibfnamefont
  {E.}~\bibnamefont {Kaxiras}},  \emph {et~al.},\ }\href@noop {} {\bibfield
  {journal} {\bibinfo  {journal} {arXiv preprint arXiv:2106.10824}\ } (\bibinfo
  {year} {2021})}\BibitemShut {NoStop}%
\bibitem [{\citenamefont {Ortiz}\ \emph {et~al.}(2019)\citenamefont {Ortiz},
  \citenamefont {Gomes}, \citenamefont {Morey}, \citenamefont {Winiarski},
  \citenamefont {Bordelon}, \citenamefont {Mangum}, \citenamefont {Oswald},
  \citenamefont {Rodriguez-Rivera}, \citenamefont {Neilson}, \citenamefont
  {Wilson} \emph {et~al.}}]{ortiz2019new}%
  \BibitemOpen
  \bibfield  {author} {\bibinfo {author} {\bibfnamefont {B.~R.}\ \bibnamefont
  {Ortiz}}, \bibinfo {author} {\bibfnamefont {L.~C.}\ \bibnamefont {Gomes}},
  \bibinfo {author} {\bibfnamefont {J.~R.}\ \bibnamefont {Morey}}, \bibinfo
  {author} {\bibfnamefont {M.}~\bibnamefont {Winiarski}}, \bibinfo {author}
  {\bibfnamefont {M.}~\bibnamefont {Bordelon}}, \bibinfo {author}
  {\bibfnamefont {J.~S.}\ \bibnamefont {Mangum}}, \bibinfo {author}
  {\bibfnamefont {I.~W.}\ \bibnamefont {Oswald}}, \bibinfo {author}
  {\bibfnamefont {J.~A.}\ \bibnamefont {Rodriguez-Rivera}}, \bibinfo {author}
  {\bibfnamefont {J.~R.}\ \bibnamefont {Neilson}}, \bibinfo {author}
  {\bibfnamefont {S.~D.}\ \bibnamefont {Wilson}},  \emph {et~al.},\ }\href@noop
  {} {\bibfield  {journal} {\bibinfo  {journal} {Phys. Rev. Mater.}\ }\textbf
  {\bibinfo {volume} {3}},\ \bibinfo {pages} {094407} (\bibinfo {year}
  {2019})}\BibitemShut {NoStop}%
\bibitem [{\citenamefont {Meier}\ \emph {et~al.}(2019)\citenamefont {Meier},
  \citenamefont {Yan}, \citenamefont {McGuire}, \citenamefont {Wang},
  \citenamefont {Christianson},\ and\ \citenamefont
  {Sales}}]{meier2019reorientation}%
  \BibitemOpen
  \bibfield  {author} {\bibinfo {author} {\bibfnamefont {W.~R.}\ \bibnamefont
  {Meier}}, \bibinfo {author} {\bibfnamefont {J.}~\bibnamefont {Yan}}, \bibinfo
  {author} {\bibfnamefont {M.~A.}\ \bibnamefont {McGuire}}, \bibinfo {author}
  {\bibfnamefont {X.}~\bibnamefont {Wang}}, \bibinfo {author} {\bibfnamefont
  {A.~D.}\ \bibnamefont {Christianson}}, \ and\ \bibinfo {author}
  {\bibfnamefont {B.~C.}\ \bibnamefont {Sales}},\ }\href@noop {} {\bibfield
  {journal} {\bibinfo  {journal} {Phys. Rev. B}\ }\textbf {\bibinfo {volume}
  {100}},\ \bibinfo {pages} {184421} (\bibinfo {year} {2019})}\BibitemShut
  {NoStop}%
\bibitem [{\citenamefont {Sales}\ \emph {et~al.}(2019)\citenamefont {Sales},
  \citenamefont {Yan}, \citenamefont {Meier}, \citenamefont {Christianson},
  \citenamefont {Okamoto},\ and\ \citenamefont
  {McGuire}}]{sales2019electronic}%
  \BibitemOpen
  \bibfield  {author} {\bibinfo {author} {\bibfnamefont {B.~C.}\ \bibnamefont
  {Sales}}, \bibinfo {author} {\bibfnamefont {J.}~\bibnamefont {Yan}}, \bibinfo
  {author} {\bibfnamefont {W.~R.}\ \bibnamefont {Meier}}, \bibinfo {author}
  {\bibfnamefont {A.~D.}\ \bibnamefont {Christianson}}, \bibinfo {author}
  {\bibfnamefont {S.}~\bibnamefont {Okamoto}}, \ and\ \bibinfo {author}
  {\bibfnamefont {M.~A.}\ \bibnamefont {McGuire}},\ }\href@noop {} {\bibfield
  {journal} {\bibinfo  {journal} {Phys. Rev. Mater.}\ }\textbf {\bibinfo
  {volume} {3}},\ \bibinfo {pages} {114203} (\bibinfo {year}
  {2019})}\BibitemShut {NoStop}%
\bibitem [{\citenamefont {Sales}\ \emph {et~al.}(2021)\citenamefont {Sales},
  \citenamefont {Meier}, \citenamefont {May}, \citenamefont {Xing},
  \citenamefont {Yan}, \citenamefont {Gao}, \citenamefont {Liu}, \citenamefont
  {Stone}, \citenamefont {Christianson}, \citenamefont {Zhang} \emph
  {et~al.}}]{sales2021tuning}%
  \BibitemOpen
  \bibfield  {author} {\bibinfo {author} {\bibfnamefont {B.}~\bibnamefont
  {Sales}}, \bibinfo {author} {\bibfnamefont {W.}~\bibnamefont {Meier}},
  \bibinfo {author} {\bibfnamefont {A.}~\bibnamefont {May}}, \bibinfo {author}
  {\bibfnamefont {J.}~\bibnamefont {Xing}}, \bibinfo {author} {\bibfnamefont
  {J.-Q.}\ \bibnamefont {Yan}}, \bibinfo {author} {\bibfnamefont
  {S.}~\bibnamefont {Gao}}, \bibinfo {author} {\bibfnamefont {Y.}~\bibnamefont
  {Liu}}, \bibinfo {author} {\bibfnamefont {M.}~\bibnamefont {Stone}}, \bibinfo
  {author} {\bibfnamefont {A.}~\bibnamefont {Christianson}}, \bibinfo {author}
  {\bibfnamefont {Q.}~\bibnamefont {Zhang}},  \emph {et~al.},\ }\href@noop {}
  {\bibfield  {journal} {\bibinfo  {journal} {Phys. Rev. Mater.}\ }\textbf
  {\bibinfo {volume} {5}},\ \bibinfo {pages} {044202} (\bibinfo {year}
  {2021})}\BibitemShut {NoStop}%
\bibitem [{\citenamefont {Kim}\ \emph {et~al.}(2017)\citenamefont {Kim},
  \citenamefont {DaSilva}, \citenamefont {Huang}, \citenamefont {Fallahazad},
  \citenamefont {Larentis}, \citenamefont {Taniguchi}, \citenamefont
  {Watanabe}, \citenamefont {LeRoy}, \citenamefont {MacDonald},\ and\
  \citenamefont {Tutuc}}]{kim2017tunable}%
  \BibitemOpen
  \bibfield  {author} {\bibinfo {author} {\bibfnamefont {K.}~\bibnamefont
  {Kim}}, \bibinfo {author} {\bibfnamefont {A.}~\bibnamefont {DaSilva}},
  \bibinfo {author} {\bibfnamefont {S.}~\bibnamefont {Huang}}, \bibinfo
  {author} {\bibfnamefont {B.}~\bibnamefont {Fallahazad}}, \bibinfo {author}
  {\bibfnamefont {S.}~\bibnamefont {Larentis}}, \bibinfo {author}
  {\bibfnamefont {T.}~\bibnamefont {Taniguchi}}, \bibinfo {author}
  {\bibfnamefont {K.}~\bibnamefont {Watanabe}}, \bibinfo {author}
  {\bibfnamefont {B.~J.}\ \bibnamefont {LeRoy}}, \bibinfo {author}
  {\bibfnamefont {A.~H.}\ \bibnamefont {MacDonald}}, \ and\ \bibinfo {author}
  {\bibfnamefont {E.}~\bibnamefont {Tutuc}},\ }\href@noop {} {\bibfield
  {journal} {\bibinfo  {journal} {Proc. Natl. Acad. Sci. U.S.A.}\ }\textbf
  {\bibinfo {volume} {114}},\ \bibinfo {pages} {3364} (\bibinfo {year}
  {2017})}\BibitemShut {NoStop}%
\bibitem [{sup()}]{supp}%
  \BibitemOpen
  \href@noop {} {\enquote {\bibinfo {title} {See supplemental material at [url
  will be inserted by publisher] for additional information about experimental
  and theoretical details and methods.}}\ }\BibitemShut {NoStop}%
\bibitem [{\citenamefont {H{\"a}ggstr{\"o}m}\ \emph
  {et~al.}(1975{\natexlab{a}})\citenamefont {H{\"a}ggstr{\"o}m}, \citenamefont
  {Ericsson}, \citenamefont {W{\"a}ppling},\ and\ \citenamefont
  {Chandra}}]{haggstrom1975studies}%
  \BibitemOpen
  \bibfield  {author} {\bibinfo {author} {\bibfnamefont {L.}~\bibnamefont
  {H{\"a}ggstr{\"o}m}}, \bibinfo {author} {\bibfnamefont {T.}~\bibnamefont
  {Ericsson}}, \bibinfo {author} {\bibfnamefont {R.}~\bibnamefont
  {W{\"a}ppling}}, \ and\ \bibinfo {author} {\bibfnamefont {K.}~\bibnamefont
  {Chandra}},\ }\href@noop {} {\bibfield  {journal} {\bibinfo  {journal} {Phys.
  Scr.}\ }\textbf {\bibinfo {volume} {11}},\ \bibinfo {pages} {47} (\bibinfo
  {year} {1975}{\natexlab{a}})}\BibitemShut {NoStop}%
\bibitem [{\citenamefont {Svanidze}\ \emph {et~al.}(2015)\citenamefont
  {Svanidze}, \citenamefont {Wang}, \citenamefont {Besara}, \citenamefont
  {Liu}, \citenamefont {Huang}, \citenamefont {Siegrist}, \citenamefont
  {Frandsen}, \citenamefont {Lynn}, \citenamefont {Nevidomskyy}, \citenamefont
  {Gam{\.z}a} \emph {et~al.}}]{svanidze2015itinerant}%
  \BibitemOpen
  \bibfield  {author} {\bibinfo {author} {\bibfnamefont {E.}~\bibnamefont
  {Svanidze}}, \bibinfo {author} {\bibfnamefont {J.~K.}\ \bibnamefont {Wang}},
  \bibinfo {author} {\bibfnamefont {T.}~\bibnamefont {Besara}}, \bibinfo
  {author} {\bibfnamefont {L.}~\bibnamefont {Liu}}, \bibinfo {author}
  {\bibfnamefont {Q.}~\bibnamefont {Huang}}, \bibinfo {author} {\bibfnamefont
  {T.}~\bibnamefont {Siegrist}}, \bibinfo {author} {\bibfnamefont
  {B.}~\bibnamefont {Frandsen}}, \bibinfo {author} {\bibfnamefont {J.~W.}\
  \bibnamefont {Lynn}}, \bibinfo {author} {\bibfnamefont {A.~H.}\ \bibnamefont
  {Nevidomskyy}}, \bibinfo {author} {\bibfnamefont {M.~B.}\ \bibnamefont
  {Gam{\.z}a}},  \emph {et~al.},\ }\href@noop {} {\bibfield  {journal}
  {\bibinfo  {journal} {Nat. Commun.}\ }\textbf {\bibinfo {volume} {6}},\
  \bibinfo {pages} {1} (\bibinfo {year} {2015})}\BibitemShut {NoStop}%
\bibitem [{\citenamefont {Matthias}\ \emph {et~al.}(1961)\citenamefont
  {Matthias}, \citenamefont {Clogston}, \citenamefont {Williams}, \citenamefont
  {Corenzwit},\ and\ \citenamefont {Sherwood}}]{matthias1961ferromagnetism}%
  \BibitemOpen
  \bibfield  {author} {\bibinfo {author} {\bibfnamefont {B.}~\bibnamefont
  {Matthias}}, \bibinfo {author} {\bibfnamefont {A.}~\bibnamefont {Clogston}},
  \bibinfo {author} {\bibfnamefont {H.}~\bibnamefont {Williams}}, \bibinfo
  {author} {\bibfnamefont {E.}~\bibnamefont {Corenzwit}}, \ and\ \bibinfo
  {author} {\bibfnamefont {R.}~\bibnamefont {Sherwood}},\ }\href@noop {}
  {\bibfield  {journal} {\bibinfo  {journal} {Phys. Rev. Lett.}\ }\textbf
  {\bibinfo {volume} {7}},\ \bibinfo {pages} {7} (\bibinfo {year}
  {1961})}\BibitemShut {NoStop}%
\bibitem [{\citenamefont {Shanavas}\ and\ \citenamefont
  {Singh}(2015)}]{shanavas2015itinerant}%
  \BibitemOpen
  \bibfield  {author} {\bibinfo {author} {\bibfnamefont {K.}~\bibnamefont
  {Shanavas}}\ and\ \bibinfo {author} {\bibfnamefont {D.~J.}\ \bibnamefont
  {Singh}},\ }\href@noop {} {\bibfield  {journal} {\bibinfo  {journal} {PloS
  one}\ }\textbf {\bibinfo {volume} {10}},\ \bibinfo {pages} {e0121186}
  (\bibinfo {year} {2015})}\BibitemShut {NoStop}%
\bibitem [{\citenamefont {H{\"a}ggstr{\"o}m}\ \emph
  {et~al.}(1975{\natexlab{b}})\citenamefont {H{\"a}ggstr{\"o}m}, \citenamefont
  {Ericsson},\ and\ \citenamefont {W{\"a}ppling}}]{haggstrom1975investigation}%
  \BibitemOpen
  \bibfield  {author} {\bibinfo {author} {\bibfnamefont {L.}~\bibnamefont
  {H{\"a}ggstr{\"o}m}}, \bibinfo {author} {\bibfnamefont {T.}~\bibnamefont
  {Ericsson}}, \ and\ \bibinfo {author} {\bibfnamefont {R.}~\bibnamefont
  {W{\"a}ppling}},\ }\href@noop {} {\bibfield  {journal} {\bibinfo  {journal}
  {Phys. Scr.}\ }\textbf {\bibinfo {volume} {11}},\ \bibinfo {pages} {94}
  (\bibinfo {year} {1975}{\natexlab{b}})}\BibitemShut {NoStop}%
\bibitem [{\citenamefont {Zhang}\ \emph {et~al.}(2021)\citenamefont {Zhang},
  \citenamefont {Yilmaz}, \citenamefont {Meier}, \citenamefont {Pai},
  \citenamefont {Lapano}, \citenamefont {Li}, \citenamefont {Kaznatcheev},
  \citenamefont {Vescovo}, \citenamefont {Huon}, \citenamefont {Brahlek} \emph
  {et~al.}}]{zhang2021flat}%
  \BibitemOpen
  \bibfield  {author} {\bibinfo {author} {\bibfnamefont {J.}~\bibnamefont
  {Zhang}}, \bibinfo {author} {\bibfnamefont {T.}~\bibnamefont {Yilmaz}},
  \bibinfo {author} {\bibfnamefont {J.}~\bibnamefont {Meier}}, \bibinfo
  {author} {\bibfnamefont {J.}~\bibnamefont {Pai}}, \bibinfo {author}
  {\bibfnamefont {J.}~\bibnamefont {Lapano}}, \bibinfo {author} {\bibfnamefont
  {H.}~\bibnamefont {Li}}, \bibinfo {author} {\bibfnamefont {K.}~\bibnamefont
  {Kaznatcheev}}, \bibinfo {author} {\bibfnamefont {E.}~\bibnamefont
  {Vescovo}}, \bibinfo {author} {\bibfnamefont {A.}~\bibnamefont {Huon}},
  \bibinfo {author} {\bibfnamefont {M.}~\bibnamefont {Brahlek}},  \emph
  {et~al.},\ }\href@noop {} {\bibfield  {journal} {\bibinfo  {journal} {arXiv
  preprint arXiv:2105.08888}\ } (\bibinfo {year} {2021})}\BibitemShut {NoStop}%
\bibitem [{\citenamefont {Matthias}\ and\ \citenamefont
  {Bozorth}(1958)}]{matthias1958ferromagnetism}%
  \BibitemOpen
  \bibfield  {author} {\bibinfo {author} {\bibfnamefont {B.}~\bibnamefont
  {Matthias}}\ and\ \bibinfo {author} {\bibfnamefont {R.}~\bibnamefont
  {Bozorth}},\ }\href@noop {} {\bibfield  {journal} {\bibinfo  {journal} {Phys.
  Rev.}\ }\textbf {\bibinfo {volume} {109}},\ \bibinfo {pages} {604} (\bibinfo
  {year} {1958})}\BibitemShut {NoStop}%
\end{thebibliography}

%


\setcounter{equation}{0}
\setcounter{figure}{0}
\setcounter{table}{0}
\setcounter{page}{1}
\makeatletter
\renewcommand{\theequation}{S\arabic{equation}}
\renewcommand{\thefigure}{S\arabic{figure}}
\renewcommand{\bibnumfmt}[1]{[S#1]}
\renewcommand{\citenumfont}[1]{S#1}

\section{Supplementary Materials}

\subsection{Sample growth, characterization, and structure}

\begin{figure*}
\begin{center}
\includegraphics[width=2.0in]{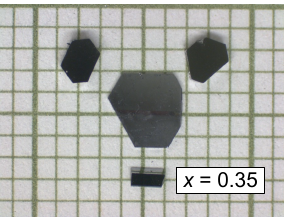}
\caption{\label{fig:supp-crystals}
Crystals from a flux growth of \cosnin\ with x = 0.35 on a mm grid.
}
\end{center}
\end{figure*}

Single crystals of \cosnin\ are grown out of a tin-indium flux, as described in detail previously \cite{sales2019electronic, sales2021tuning, meier2020flat, meier2019reorientation}. Crystals with a composition of CoSn$_{0.73}$In$_{0.27}$ are grown from a starting composition of 25 g In (99.999), 5 g Sn (99.9999) and 0.5 g Co (99.95) in a 10 cc alumina crucible sealed inside an evacuated silica ampoule. After homogenizing the liquid at 1130\degreesC\ for 24 h, and physically shaking the liquid at 900\degreesC, the melt is cooled at 1\degreesC/h to 620\degreesC. At 620\degreesC\ the silica ampoule is removed from the furnace and the excess Sn-In flux is centrifuged into a 10 cc catch crucible filled with coarse porosity quartz wool. For In concentrations below x$\approx$0.3 the crystals grow as hexagonal bars with typical dimensions of a few mm on a side with lengths up to 20 mm. For values of x $>$ 0.3, the crystal morphology changes to primarily hexagonal plates a few mm on a side and less than 1 mm thick and the yield of crystals drops dramatically. For example, from approximately 25 grams of an indium rich flux, only about 50-100 mg of crystals of composition CoSn$_{0.6}$In$_{0.4}$ precipitate from the melt with each crystal weighing less than 1 mg. Crystals from a CoSn$_{0.65}$In$_{0.35}$ growth are shown in Fig. \ref{fig:supp-crystals}. The chemical composition of several crystals from each growth are checked using an Hitachi TM-3000 electron microscope equipped with a Bruker Quantax 70 for energy dispersive spectroscopy (EDS). All of the In-doped crystals are chemically homogenous within the limits of this analysis. The EDS determined composition is used to label the crystals in this work.

Polycrystalline samples of \cosnin\ are prepared by using $\approx$1\,mm pieces of Co metal, In shot and Sn shot. Stoichiometric amounts of each element with a typical total mass of 5 g are loaded into an evacuated clean round bottom silica ampoule. The ampoule is heated to 600\degreesC\ for a few hours and then heated to 1100\degreesC\ for 20 h, followed by 1180\degreesC\ for 2-4 h. The ampoule is removed from the furnace at 1180\degreesC\ and quenched into water, making sure that the liquid remains at the round bottom of the sealed ampoule. After cooling to room temperature, the ampoule and sample are annealed in a furnace for 4-5 days at 800\degreesC. The phase purity and lattice constants from both polycrystalline and single crystals of \cosnin\ are determined from powder x-ray diffraction using a PANalytical X'pert Pro diffractometer with Cu $K_{\alpha1}$ radiation. No significant mass loss was observed in the polycrystalline growths, and no significant secondary phases were observed by x-ray diffraction. EDS measurements on powder grains gave values of x within 0.01-0.02 of the nominal (loaded) compositions.  The nominal composition is used to label the polycrystalline samples in this work.

Magnetic susceptibility, resistivity and specific heat measurements are made using commercial equipment from Quantum Design. Powder neutron diffraction data is collected at both the Spallation Neutron Source SNS (POWGEN) and the High flux Isotope Reactor, HFIR, beamlines HB-2A, and HB-1B. Tin-119 M\"{o}ssbauer spectra were recorded in a constant acceleration mode spectrometer in the $\pm$10 mm/s velocity range at room temperature using a Ca$^{119m}$SnO$_3$ source with a Wissel drive and CMCA-550 data acquisition unit. To reduce Sn fluorescence from the source and sample, an indium foil was placed between the powdered samples mixed with boron-nitride and a sodium iodide scintillator.

\begin{figure*}
\begin{center}
\includegraphics[width=6.0in]{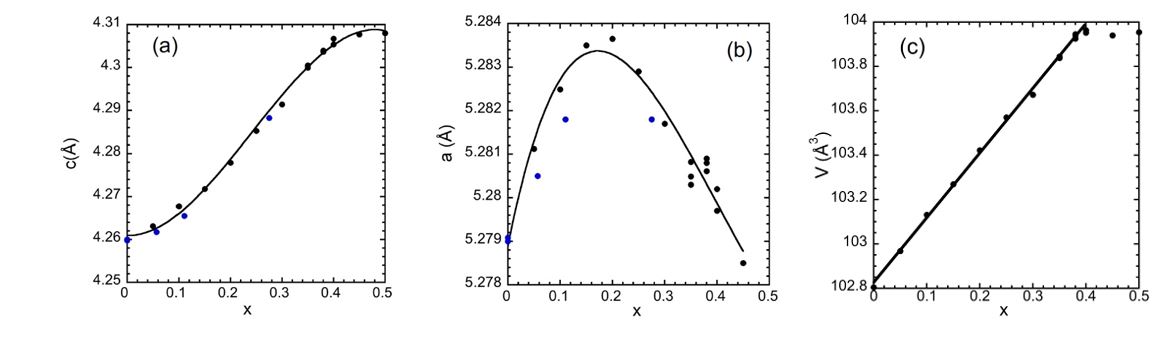}
\caption{\label{fig:supp-lattice}
Variation of the lattice constants and unit cell volume of \cosnin\ with x. The black circles are from polycrystalline samples (x corresponding to the loaded Sn/In ratio) and the blue circles from single crystals (x measured by EDS). The weak variation of c and V with x for values of x $>$ 0.4 signals the solubility limit for In in the CoSn structure. The unusual variation of the a lattice constant with x reflects preferential substitution of In into one of the two Sn sites in the CoSn structure (see text for more details).
}
\end{center}
\end{figure*}

The variation of the lattice parameters and volume with indium concentration, x, are shown in Fig. \ref{fig:supp-lattice}. The c lattice parameter and volume increase up to about x = 0.40, which is the solubility limit for indium in both single crystals and polycrystalline samples prepared as described. The percentage variation of the a lattice constant with doping is smaller than c, but it is clearly non-monotonic and peaks near x $\approx$ 0.2. A likely explanation is that the In dopant prefers one of the two Sn sites in the CoSn structure. A site preference could not be determined using x-ray diffraction because of scattering factors for Sn and In. The site preference could be determined using powder neutron diffraction and Sn M\"{o}ssbauer measurements, as shown below.

\subsection{Neutron diffraction}

\begin{figure*}
\begin{center}
\includegraphics[width=4.0 in]{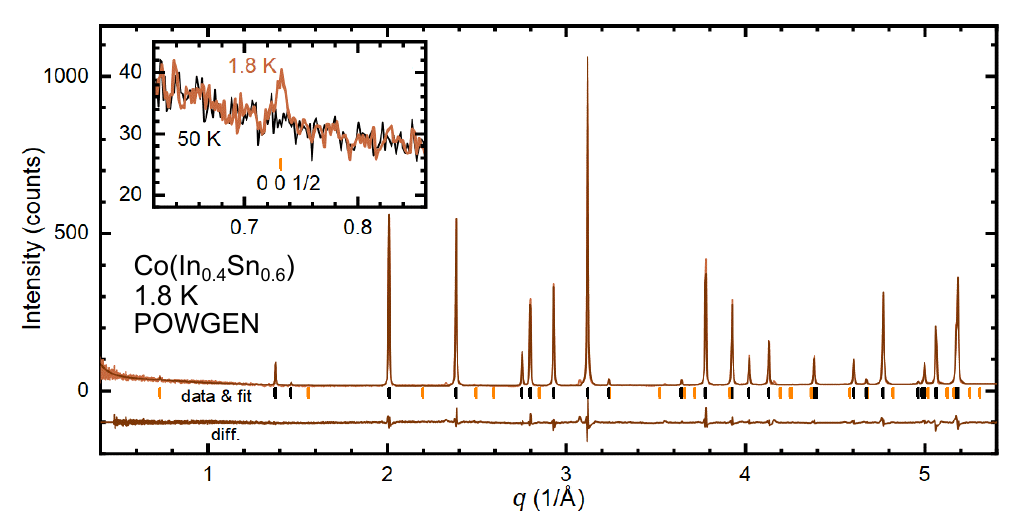}
\caption{\label{fig:supp-powgen}
Rietveld refinement of powder neutron diffraction data collected at 1.8 K. Inset shows the region near the 0 0 $\frac{1}{2}$ magnetic peak at 1.8 and 50\,K.
}
\end{center}
\end{figure*}

A full neutron diffraction pattern collected at POWGEN at 1.8\,K for CoSn$_{0.6}$In$_{0.4}$ is shown in Fig. \ref{fig:supp-powgen}. The magnetic peak at 0 0 $\frac{1}{2}$ is shown in the inset. The magnetic space group for the model presented in the main text is determined to be $C_cmcm$ (\#63.466). This is consistent with A-type magnetic order; however, we cannot completely exclude another possible magnetic structure, i.e., the in-plane noncollinear spin ordering with antiferromagnetic arrangement along the out-of-plane direction.

\subsection{Indium site preference}

To examine the origin of the non-linear evolution of the lattice constants of \cosnin\ with increasing x (Fig. \ref{fig:supp-lattice}) we set out to determine the site occupation of the Sn sites by M\"{o}ssbauer spectroscopy and neutron diffraction. CoSn has two tin sites. Sn1 (Wyckoff 1a) sits within the hexagons of the Co-kagome layer and Sn2 (Wyckoff 2d) forms a honeycomb network between the kagome sheets. Critically, we would expect different lattice evolution by doping the larger indium atoms onto one site or another. For example, stuffing In into the 1a-site within the close-packed SnCo$_3$ plane might expand the a lattice parameter more than c. In an extreme case, In preferentially filling this site first, we would expect the lattice to expand quickly in plane until the site is completely filled (around x = 1/3). For larger dopings, indium would have to start filling the honeycomb 2d site expanding the c axis faster. This is roughly what is observed in the lattice parameter evolutions presented in Fig. \ref{fig:supp-lattice}.

\begin{figure}
\begin{center}
\includegraphics[width=2.75in]{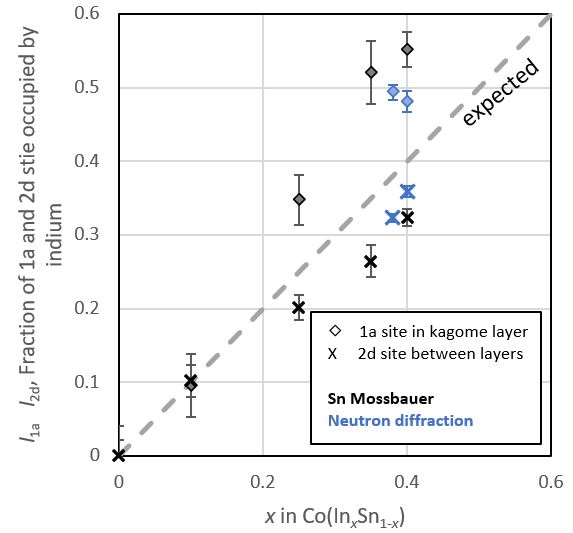}
\caption{\label{fig:supp-sitepref}
Occupancy of the two Sn sites as the concentration of indium, x, is changed in the CoSn structure. Blue points are from powder neutron diffraction data, and the black points are from M\"{o}ssbauer data. For x near 0.4, In prefers the 1a site in the kagome layer by about a factor of 2 to 1.
}
\end{center}
\end{figure}

To test this idea, we sought indium site occupations on the 1a and 2d sites across the \cosnin\ series. Indium and tin are adjacent, heavy elements and therefore there is little x-ray scattering contrast between them. Fortunately, they have moderate neutron scattering contrast. We obtained the indium fraction on the two sites, defined here as I1a and I2d, by refining our HB-2A and POWGEN neutron diffraction data from x = 0.38 and x = 0.4 powders. To do this we forced the indium fractions to give the correct overall composition; x = (I1a + 2 I2d)/3. As shown in Figure \ref{fig:supp-sitepref}, the resulting values show a 50\% and 30\% higher fractional indium occupation of the 1a than 2d sites. There is a clear site preference for In on the 1a site within the kagome plane.

M\"{o}ssbauer spectroscopy gives us an alternative method to extract the fractional occupations of the two sites. The two sites in CoSn have distinctly different M\"{o}ssbauer parameters \cite{haggstrom1975investigation} allowing us to estimate the relative number of Sn on each site. The M\"{o}ssbauer spectra of \cosnin\ are similarly comprised of two doublets that correspond to the Sn1 and Sn2 sites of the structure. According to the site multiplicity, a spectral area ratio of Sn1/Sn2 of 1:2 is expected, if the Lamb-M\"{o}ssbauer factor for Sn at both sites is considered equal, as is observed in Ref. \cite{haggstrom1975investigation} and in our data for x = 0. The isomer shift and quadrupole splitting for both sites are found to be consistent with those of intermetallic Sn and those in Ref. \cite{haggstrom1975investigation}. Examining the data for samples with x$>$0, we do not observe any significant change in the spectral parameters as a function of indium content. In contrast, the ratio of Sn1/Sn2 spectral area increases from 1:2 (1/3 on the 1a site and 2/3 on the 2d site) for x=0, to 1:3 (1/4 on the 1a site and 3/4 on the 2d site), for x=0.4, which corresponds to a small but significant preference for In to occupy the Sn1 site and indicates that the site occupation of indium does not occur randomly. For CoSn we expect to find the fraction of all tin atoms on the 1a site, F1a, to be 1/3 because there are two 2d sites and one 1a site per unit cell. If indium has no site preference, it will occupy these sites so that F1a will remain 1/3. This is not what we observe. We calculate the fraction of 1a sites that have indium using the overall composition.

\begin{equation*}
  I1a = 1 – 3 F1a (1 - x) ;	I2d = 1/2 (3 x – 1 + 3 F1a (1 – x))
\end{equation*}

Figure \ref{fig:supp-sitepref} clearly shows a consistent preference for indium on the 1a site especially for larger indium concentrations. Both neutron diffraction and M\"{o}ssbauer spectroscopy reveal indium's site preference for the 1a site within the Co kagome layer. The evolution of the site fillings by indium is likely the origin of the non-linear trends of the lattice parameters across the \cosnin\ series (see Fig. \ref{fig:supp-lattice}).

\subsection{Angle Resolved Photoemission Spectroscopy}

\begin{figure*}
\begin{center}
\includegraphics[width=4.0in]{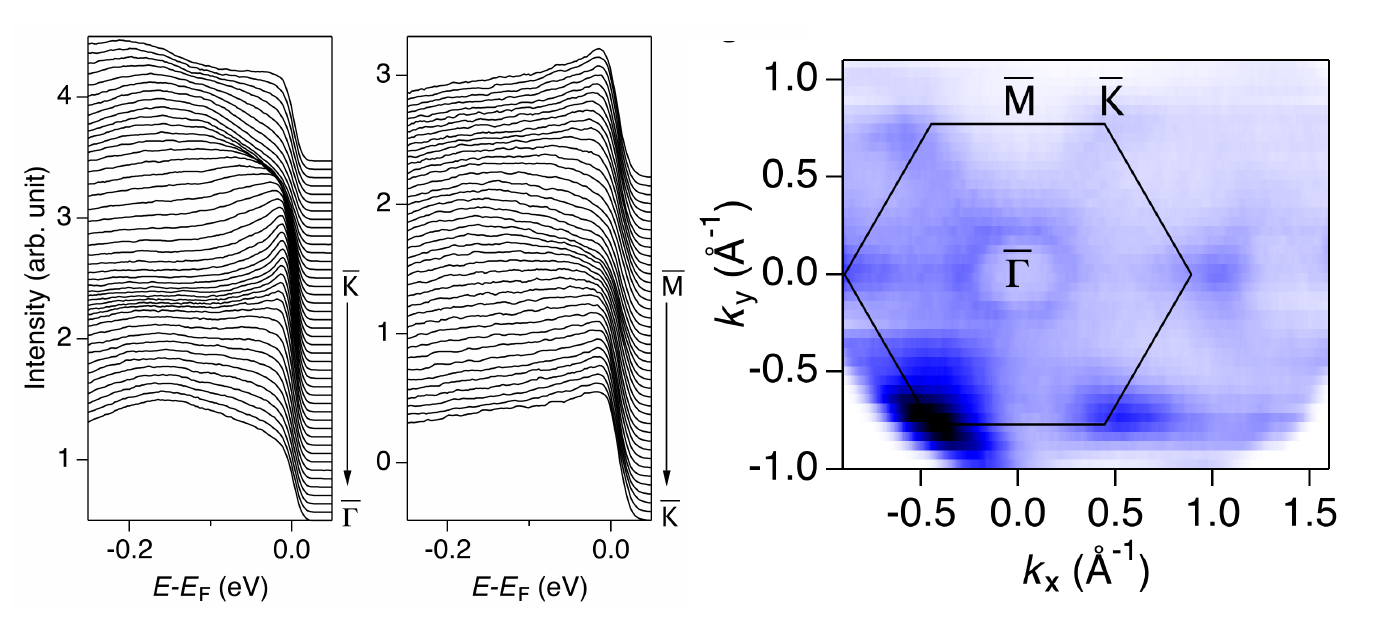}
\caption{\label{fig:supp-ARPES}
Additional ARPES results. Second derivative of EDC curves near E$_F$ along different crystallographic directions and a Fermi surface map for CoSn$_{0.6}$In$_{0.4}$.
}
\end{center}
\end{figure*}

The ARPES experiments were performed at beamline 21-ID-1 at National Synchrotron Light Source II, Brookhaven. All samples were cleaved in-situ in a vacuum better than $5\times10^{-11}$ Torr. The measurements were taken with synchrotron light source and a Scienta-Omicron DA30 electron analyzer. The total energy resolution was set to be 15 meV. The sample stage was maintained at low temperature (T=15 K) throughout the experiment. Additional EDC curves and a Fermi surface map, collected along with the data shown in the main text, are shown in Fig. \ref{fig:supp-ARPES}

\subsection{Electrical resistivity}

Resistivity measurements were made on small rectangular bar shaped crystals with silver epoxy (H20E) electrical contacts and 0.001 or 0.002 inch diameter platinum wire. Before the measurements, a pulsed DC current (30 mA) was used to lower the contact resistance to less than 1 $\Omega$. The long dimension for some of the crystals was slightly less than 1 mm resulting in about a 50\% uncertainty in the absolute value of the resistivity.

\begin{figure}
\begin{center}
\includegraphics[width=2.25in]{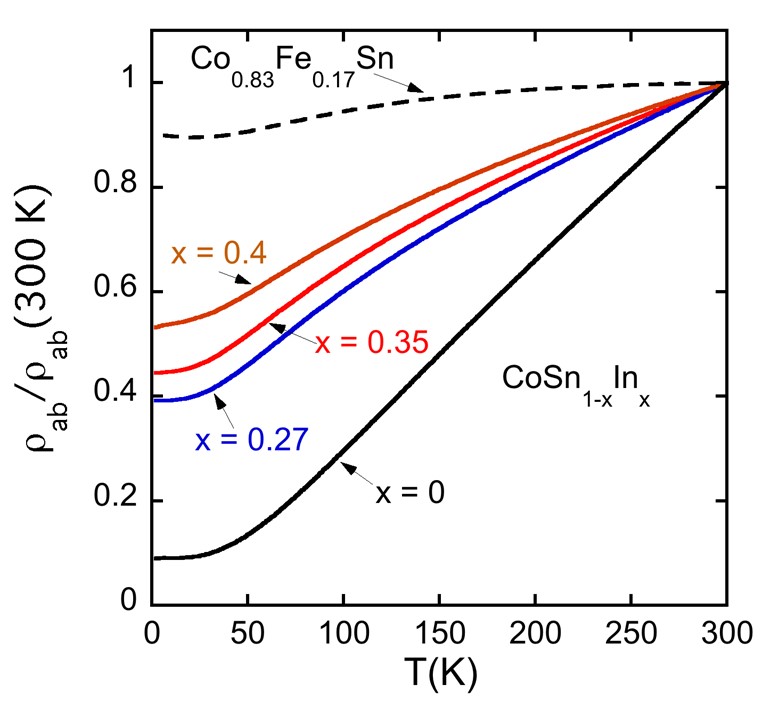}
\caption{\label{fig:supp-res}
Resistivity from single crystals of \cosnin\ for x =0, 0.27, 0.35, 0.40. For all of the crystals the resistivity in the kagome planes ($\rho_{ab}$) is 10 to 20 times larger than $\rho_{c}$ \cite{sales2019electronic, sales2021tuning, meier2020flat, meier2019reorientation}. With increasing In content there is a remarkable increase in the magnitude of the resistivity associated with the movement of the Fermi energy into the flat bands and the onset of magnetic order. The room temperature values of $\rho_{ab}$ for the x = 0.27, 0.35 and 0.4 crystals are all about 300 $\mu\Omega$cm. For x =0.40 the magnetic order at 30 K is not apparent in the resistivity data or its temperature derivative. For comparison, $\rho_{ab}$ from an iron-doped CoSn crystal (Co$_{0.83}$Fe$_{0.17}$Sn) with a similar room temperature resistivity is shown by the dashed line. The iron-doped material undergoes a spin-glass transition at 6 K \cite{sales2021tuning} which also is not evident in the electrical transport data.
}
\end{center}
\end{figure}

The in-plane electrical resistivity $\rho_{ab}$ versus temperature for several In-doped CoSn crystals and one Fe-doped crystal with composition Co$_{0.83}$Fe$_{0.17}$Sn is shown in Fig. \ref{fig:supp-res}. Each curve is normalized to 1 at 300 K . All of the crystals are good metals with the room temperature resistivity reaching a value of about 300 $\mu\Omega$cm for the more heavily doped samples. Note the much larger effect Fe doping has on the resistivity curves. The crystals with 0.35 and 0.4 concentrations of In have resistivities that are nearly linear at the lowest temperatures (2-10 K), indicating substantial non-Fermi liquid behavior. All of the magnetically ordered crystals have a negative transverse magnetoresistance of a few percent at 2 K (also see Ref. \citenum{zhang2021flat}).

\subsection{Heat capacity}

\begin{figure*}
\begin{center}
\includegraphics[width=6.0 in]{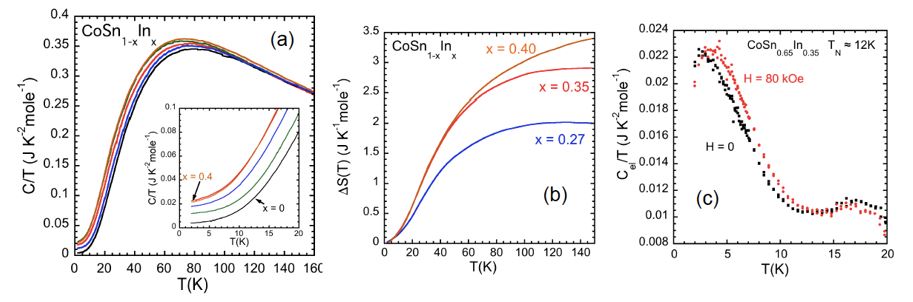}
\caption{\label{fig:supp-hc}
Heat capacity divided by temperature for \cosnin\ crystals with x = 0, 0.11, 0.27, 0.35 and 0.4. For temperatures above about 160 K the C/T data converge for all values of x. (b) Excess entropy for samples with x = 0.27, x = 0.35 and x = 0.4 relative to pure CoSn. Note that on cooling entropy is removed beginning near 150 K. For the x = 0.40 sample the total change in entropy relative to CoSn is about 0.6R$ln$(2). (c) Estimated electronic/magnetic heat capacity for a CoSn$_{0.65}$In$_{0.35}$ crystal with x = 0.35, and $T_N \approx$ 10 K after subtracting the lattice contribution. See text for details.
}
\end{center}
\end{figure*}

The low temperature specific heat data from a series of indium-doped CoSn crystals are shown in Fig \ref{fig:supp-hc}a. There is no clear evidence of a phase transition in the raw heat capacity data. This is not surprising because of the small magnitude of the ordered magnetic moment and the highly itinerant nature of the magnetic order. In weak itinerant ferromagnets such as \ce{ZrZn2}, and \ce{Sc3In}, the entropy released near T$_C$ is typically about 1-3\% of the entropy expected for a mole of S = 1/2 spins (R$ln$(2)= 5.76 J K$^{-1}$mole$^{-1}$) \cite{matthias1958ferromagnetism, matthias1961ferromagnetism}. Weak itinerant antiferromagnets appear to be quite rare. One example is orthorhombic TiAu, which has T$_N$ = 36 K and an ordered moment of 0.15 $\mu_B$ per Ti \cite{svanidze2015itinerant} with an entropy release near T$_N$ of 3\% of R$ln$(2).

Indium doped crystals with x = 0.35 have a small feature in the magnetic susceptibility near 8 K, which we attribute to antiferromagnetic order. For this composition, plots of C/T vs T$^2$ are extremely linear from 12 to 20 K, which suggests that data from this temperature interval can be used to accurately extract the lattice contribution to the specific heat data for T $<$ 20 K. After subtracting the lattice contribution, the remaining electronic/magnetic contribution to C/T is shown in Fig. \ref{fig:supp-hc}c. While there is some entropy released on heating to about 15 K, it only amounts to about 1.5\% of R$ln$(2). The effect of an 80 kOe magnetic field on the peak in Fig. \ref{fig:supp-hc}c is minimal, presumably because of the small magnetic moments. This analysis, however, makes several assumptions and should only be regarded as qualitative.

A simpler, more quantitative analysis of the specific heat data simply compares the specific heat of the indium-doped crystals to that of CoSn. As is evident in Fig. \ref{fig:supp-hc}a, with increasing indium content the value of the low temperature Sommerfeld coefficient, $\gamma$, increases from 3.7 mJ K$^{-2}$mole$^{-1}$ for CoSn to 22 mJ K$^{-2}$mole$^{-1}$ for CoSn$_{0.6}$In$_{0.4}$ (see main text). The excess specific heat for the In-doped crystals relative to pure CoSn persists up to 100-160 K. Above 160 K the specific heat data from all of the indium-doped crystals are the same as for pure CoSn, within experimental error. Integration of the excess in C/T for the indium-doped crystals relative to the C/T data from CoSn yields the entropy evolution on heating to 160 K of 30 - 60\% of R$ln$(2) (Fig. \ref{fig:supp-hc}b). This analysis is plausible, particularly in a quasi-2D material, since the very electrons that generate a high density of states are the same electrons that generate the weak itinerant antiferromagnetism. For example, CoSn$_{0.6}$In$_{0.4}$ crystals have T$_N$ =30 K, but about 75\% of the entropy is removed above T$_N$, which is typical for a quasi-2D magnet. In the case of a robust quasi-2D magnet with more localized magnetism like FeSn only about 10\% of the entropy is removed near T$_N$ and most is removed well above T$_N$ as short-range magnetic order in each kagome layer \cite{sales2019electronic}.

\subsection{DFT calculations}

First principles calculations were carried out within the Generalized Gradient Approximation, using the linearized augmented plane-wave density functional theory code WIEN2K. Sphere radii of 2.46, 2.48 and 2.46 Bohr were used, respectively for Co, Sn and In, and an RK$_{max}$ value of 9.0 (the product of the smallest sphere radius and the largest plane-wave expansion vector) was employed; spin-orbit coupling was not included. Lattice parameters of CoSn were used for all calculations. Relative energies and ordered moments from calculations using three different approaches are collected in Table \ref{table-DFT}, where NM means non-magnetic, FM means ferromagnetic, and AF means antiferromagnetic with FM layers stacked AF.

\begin{table*}
\caption{First principles-calculated properties of the CoSn$_{2/3}$In$_{1/3}$ material in three approximations. ``Substitution'' -- the onefold Sn site is replaced in the unit cell by In. ``VCA'' -- the virtual crystal approximation, where all Sn atoms are assumed to have ionic charge Z= 49.66667, with the total number of electrons in the cell adjusted accordingly; ``+U'' -- the same unit cell as in the ``Substitution'' calculation is used, with a U of 2 eV applied to the Co d orbitals (self-interaction-corrected option). The Moment column lists the total moment within the Co sphere. In the Energy column, “0” represents the ground state. }
\begin{tabular}{l|c|c|c}
\hline
Approach 	&	Magnetic state	&	Moment ($\mu_B$ / Co)	&	Relative energy (meV / Co)	\\
\hline
	&	NM	&	--	&	17.18	\\
GGA, Substitution	&	FM	&	0.419	&	0	\\
	&	AF	&	0.429	&	0.51	\\
\hline
	&	NM	&	--	&	9.66	\\
GGA, VCA	&	FM	&	0.274	&	0	\\
	&	AF	&	0.277	&	3.24	\\
\hline
	&	NM	&	--	&	160.41	\\
GGA+U, Substitution	&	FM	&	1.100	&	11.05	\\
	&	AF	&	1.156	&	0	\\
\hline
\end{tabular}\
\label{table-DFT}
\end{table*}

\end{document}